\documentclass{nature}

\usepackage{caption}
\usepackage{graphicx}
\usepackage[scaled]{helvet}
\usepackage{amsmath}
\usepackage{amssymb}
\usepackage{bm}
\usepackage{xfrac}
\newcommand*{\myfont}{\fontfamily{phv}\selectfont}
\usepackage{hyperref}
\usepackage{anyfontsize}
\usepackage{ragged2e}

\usepackage{lineno}
\usepackage{color, colortbl}

\definecolor{CLBlue}{rgb}{0, .25, .8}
\definecolor{MyGreen}{rgb}{.08, .47, .40}
\definecolor{MyMagenta}{rgb}{.59, .08, .35}
\definecolor{MyOrange}{rgb}{.78, .31, 0}
\definecolor{MyGrey}{rgb}{.5, .5, .5}

\title{Emergent scale--free networks}

\author{Christopher W. Lynn$^{1,2}$, Caroline M. Holmes$^2$, \& Stephanie E. Palmer$^{3,4}$}

\begin{document}

%\linenumbers

\maketitle

\begin{affiliations}
\item Initiative for the Theoretical Sciences, Graduate Center, City University of New York, New York, NY 10016, USA
\item Joseph Henry Laboratories of Physics and Lewis--Sigler Institute for Integrative Genomics, Princeton University, Princeton, NJ 08544, USA
\item Department of Organismal Biology and Anatomy, University of Chicago, Chicago, IL 60637, USA
\item Department of Physics, University of Chicago, Chicago, IL 60637, USA
\end{affiliations}

%\newpage

\section*{\large Abstract}
\vspace{-30pt}
\noindent\rule{\textwidth}{.5pt}

\begin{abstract}

Many complex systems---from social and communication networks to biological networks and the Internet---are thought to exhibit scale--free structure. However, prevailing explanations rely on the constant addition of new nodes, an assumption that fails dramatically in some real--world settings. Here, we propose a model in which nodes are allowed to die, and their connections rearrange under a mixture of preferential and random attachment. Under these simple dynamics, we show that networks self--organize towards scale--free structure, with a power--law exponent $\gamma = 1 + \frac{1}{p}$ that depends only on the proportion $p$ of preferential (rather than random) attachment. Applying our model to several real networks, we infer $p$ directly from data, and predict the relationship between network size and degree heterogeneity. Together, these results establish that realistic scale--free structure can emerge naturally in networks of constant size and density, with broad implications for the structure and function of complex systems.

\end{abstract}

\newpage

\section*{\large Introduction}
\vspace{-30pt}
\noindent\rule{\textwidth}{.5pt}

\noindent Scale--free structure is a hallmark feature of many complex networks, with the probability of a node having $k$ links (or degree $k$) following a power law $k^{-\gamma}$. First studied in networks of scientific citations,\cite{Price-01, Price-02} scale--free structure has now been reported across a staggering array of complex systems, from social networks (of romantic relationships,\cite{Liljeros-01} scientific collaborations,\cite{Redner-01} and online friendships\cite{Adamic-01}); to biological networks (of connections in the brain,\cite{Eguiluz-01} metabolic interactions,\cite{Albert-02} and food webs\cite{Dunne-01}); to the online and physical wiring of the Internet;\cite{Albert-03, Huberman-01, Adamic-02, Yook-01} to language,\cite{Steyvers-01} transportation,\cite{Verma-01} and communication networks.\cite{Ebel-01} Although empirically measuring power laws in real networks poses important technical challenges,\cite{Clauset-01, Broido-01} the study of scale--free structure continues to provide deep insights into the nature of complex systems.

Scale--free networks are highly heterogeneous (or heavy--tailed), with a small number of well--connected hub nodes dominating in a sea of low--degree nodes.\cite{Barabasi-03} This heterogeneity has critical implications for the function and dynamics of such systems.\cite{Albert-01} In networks of Kuramoto oscillators, for example, the transition to synchronization depends precisely on the power--law exponent $\gamma$.\cite{Rodrigues-01} Similarly, in Ising models, the critical temperature defining the phase transition from disorder to order varies systematically with $\gamma$.\cite{Dorogovtsev-01, Bianconi-01} Scale--free structure has also been used to explain the spread of viruses,\cite{Pastor-01} to analyze the robustness of complex systems to errors and attacks,\cite{Albert-04} and to investigate the communication efficiency and compressibility of information networks.\cite{Lynn-07, Lynn-10, Lynn-08}

Despite extensive investigations, there remains a basic limitation in our understanding of how scale--free structure emerges in real systems. Prevailing explanations primarily rely on two mechanisms: growth (wherein nodes are constantly added to the network) and preferential attachment (such that well--connected nodes are more likely to gain new connections).\cite{Price-02, Barabasi-03} While alternatives have been proposed to preferential attachment (such as random attachment to edges,\cite{Dorogovtsev-02} random copying of neighbors,\cite{Kumar-01} and deterministic attachment rules\cite{Barabasi-01}), the dependence on growth remains widespread.\cite{Albert-01, Moore-01, Pachon-01} In many real--world contexts, however, this dependence on constant growth is unrealistic.\cite{Park-01, Moore-01, Xie-01} In biological networks, for example, brains do not grow without bound,\cite{Lynn-05} and just as animals or species are added to a population, others die out.\cite{Dunne-01, Dunne-02} In these systems, rather than relying on growth, scale--free structure emerges organically at relatively constant size.

To describe such systems, a number of models have been proposed for scale--free networks without growth.\cite{Park-01, Xie-01, Carlson-01, Li-01, Caldarelli-02, Garlaschelli-01} For instance, power--law degree distributions can result from the optimization of network properties or by connecting nodes based on fitness.\cite{Carlson-01, Li-01, Caldarelli-02, Garlaschelli-01} However, these explanations rely on global choices for the optimization or fitness functions, and therefore do not address the self--organization of network structure. Meanwhile, there exist models for the self--organization of power--law degree distributions,\cite{Park-01, Xie-01} but these yield unrealistic exponents $\gamma \le 1$, whereas most real--world exponents lie in the range $2 \le \gamma \le 3$. Thus, understanding whether, and how, realistic scale--free structure self--organizes remains a central open question.

Here, we begin by analyzing the dynamics of real networks, demonstrating empirically that systems can maintain scale--free structure even without growth. To explain this observation, we propose an intuitive model in which nodes die at random, and the disconnected edges reattach to new nodes either preferentially (with probability $p$) or randomly (with probability $1-p$). Under these simple dynamics, the number of edges is held constant, and the network quickly approaches a steady--state size. Importantly, we show (both analytically and numerically) that scale--free structure emerges naturally, with a realistic power--law exponent $\gamma = 1 + \frac{1}{p} \ge 2$ that depends only on the proportion $p$ of preferential attachment.
 
 \section*{\large Results}
\vspace{-30pt}
\noindent\rule{\textwidth}{.5pt}

\noindent {\large \myfont Emergent scale--free structure in real networks}

\noindent In some complex systems---including many biological, language, and real--life social networks---topological properties (such as scale--free structure) arise without constant growth.\cite{Park-01, Xie-01} Meanwhile, other systems---particularly online social and communication networks, scientific collaborations, and the Internet---are often viewed as growing by accumulating new nodes and edges over time.\cite{Albert-01} Yet even for these networks, we will see that scale--free structure can arise without growth.

\begin{figure}[t]
\centering
\includegraphics[width = \textwidth]{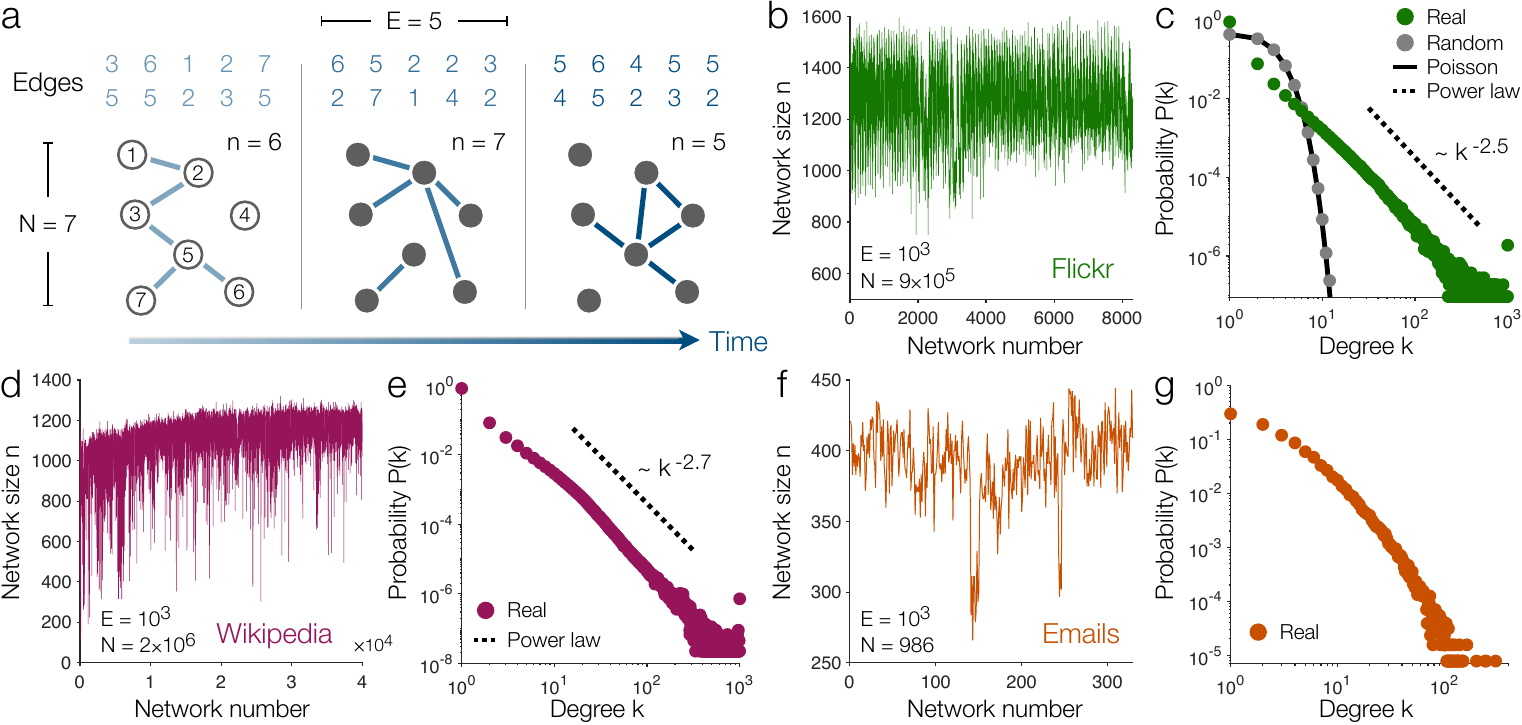} \\
\raggedright
\captionsetup{labelformat=empty}
{\spacing{1.25} \caption{\small \textbf{Fig.~\ref{fig_real} $|$ Degree distributions of real dynamical networks.} \textbf{a}, Procedure for measuring network dynamics. We divide the sequence of edges into groups of equal size $E$, thus forming a series of network snapshots. Each snapshot contains $n \le N$ nodes, where $N$ is the total number of nodes in the dataset. \textbf{b}, Trajectory of system size $n$ over time for the network of friendships on Flickr.\cite{Mislove-02} \textbf{c}, Degree distribution of the Flickr network (green) computed across all network snapshots. Randomizing each snapshot (grey) yields a Poisson distribution (solid line). Dashed line illustrates a power law for comparison. \textbf{d-e}, Trajectory of network size (\textbf{d}) and degree distribution (\textbf{e}) for the hyperlinks between pages on English Wikipedia.\cite{Mislove-01} \textbf{f-g}, Trajectory of network size (\textbf{f}) and degree distribution (\textbf{g}) for emails between scientists at a research institution.\cite{Paranjape-01} \label{fig_real}}}
\end{figure}

The dynamics of a network are defined by a sequence of connections $(i_t, j_t)$, ordered by the time $t$ at which they occur. Letting these edges accumulate over time, we arrive at a single growing network. Alternatively, one can divide the connections into groups of equal size $E$, thus defining a sequence of independent snapshots, each representing the structure of the network within a specific window of time (Fig. \ref{fig_real}a). For clarity, we let $N$ denote the total number of nodes in the sequence, while $n$ reflects the size of a single snapshot (Fig. \ref{fig_real}a). Consider, for example, the social network of friendships on Flickr (Fig. \ref{fig_real}b,c).\cite{Mislove-02} Dividing the sequence of connections into groups of size $E = 10^3$, we can study the evolution of different network properties. In particular, we find that the Flickr network fluctuates around a constant size $n$ (Fig. \ref{fig_real}b). Yet even without growing, we see that the network maintains a clear power--law degree distribution (Fig. \ref{fig_real}c), and we verify that this scale--free structure remains consistent over time (see Supplementary Information). By contrast, if we randomize the edges in each snapshot, then the degrees drop off super--exponentially as a Poisson distribution, and the scale--free structure vanishes (Fig. \ref{fig_real}c).

We can repeat the above procedure for any time--evolving network, such as links between pages on Wikipedia (Fig. \ref{fig_real}d,e) or email correspondence among scientists (Fig. \ref{fig_real}f,g).\cite{Mislove-01, Paranjape-01} Across a number of different social, web, communication, and transportation networks (see Table \ref{tab} and Methods for details on network selection), we divide the dynamics into snapshots with $E = 10^3$ edges each, the largest number that can be applied to all systems. While some networks grow slowly in time (such as Wikipedia in Fig. \ref{fig_real}d), all of the networks approach a steady--state size (see Supplementary Information). In fact, the snapshots are limited to $n \le 2E$ by definition, and therefore cannot grow without bound. Even still, many of the networks exhibit scale--free structure (such as Wikipedia in Fig. \ref{fig_real}e). We note that some of the networks are not scale--free (such as the emails in Fig. \ref{fig_real}g), but even these still display heavy--tailed degree distributions with many of the same structural properties.\cite{Stumpf-01} In what follows, we will develop a simple dynamical model capable of describing all of these networks.

\noindent {\large \myfont Model of emergent scale--free networks}

\begin{figure}[t!]
\centering
\includegraphics[width = .95\textwidth]{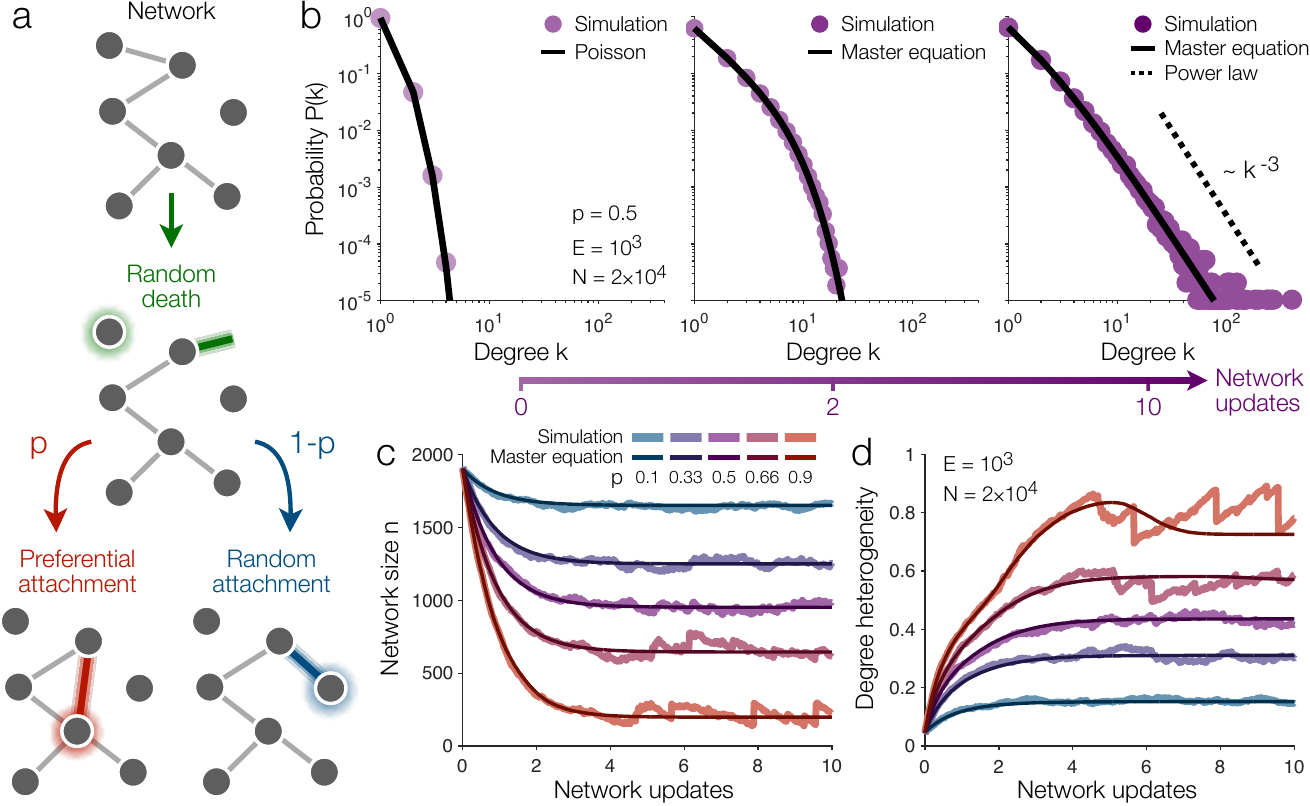} \\
\raggedright
\captionsetup{labelformat=empty}
{\spacing{1.25} \caption{\small \textbf{Fig.~\ref{fig_model} $|$ Modeling the emergence of scale--free structure.} \textbf{a}, Illustration of network dynamics. Beginning with an arbitrary network (\textit{top}), at each step in time a random node dies, losing all of its connections (\textit{center}). Each disconnected edge attaches to another node either preferentially (with probability $p$; \textit{bottom left}) or randomly (with probability $1-p$; \textit{bottom right}). \textbf{b}, Degree distributions for initially random networks (\textit{left}), after two full network updates (that is, after $2N$ steps of the dynamics; \textit{center}), and after ten network updates (\textit{right}). Distributions are computed over 100 simulations, each containing $E = 10^3$ edges, $N = 2\times 10^4$ nodes (for average degree $\bar{k} = 0.1$), and equal amounts of preferential and random attachment ($p = 0.5$). Solid lines depict predictions from the master equation [Eq. (\ref{eq_master})], and dashed line illustrates a power law for comparison. \textbf{c-d}, Trajectories of network size $n$ (\textbf{c}) and degree heterogeneity (\textbf{d}) over the course of ten network updates ($10N$ steps of the dynamics) for different values of $p$. Thick lines reflect individual simulations (beginning with random networks), and thin lines represent master equation predictions. See Methods for a detailed description of simulations. \label{fig_model}}}
\end{figure}

\noindent The above results demonstrate that scale--free structure can arise without growth in real networks. But how can we explain this observation? Here we present a simple model in which scale--free structure emerges through self-organization, with connections rearranging under a mixture of preferential and random attachment. We begin with an arbitrary network of $N$ nodes and $E$ edges (for simplicity, we always begin with a random network). At each time step, one node dies at random, losing all of its connections (Fig. \ref{fig_model}a, \textit{center}). Each of these connections then reattaches in one of two ways: (i) with probability $p$, it connects to a node via preferential attachment (that is, it attaches to node $i$ with probability proportional to its degree $k_i$; Fig. \ref{fig_model}a, \textit{bottom left}), or (ii) with probability $1-p$, it connects to a random node (Fig. \ref{fig_model}a, \textit{bottom right}). In this way, the total numbers of nodes $N$ and edges $E$ remain constant, with the wiring between nodes simply rearranging over time. Notably, besides $N$ and $E$, the model only contains a single parameter $p$, representing the proportion of preferential (rather than random) attachment.

Do the above dynamics produce scale--free structure? To answer this question, we can write down a master equation describing the evolution of the degree distribution $P_t(k)$ from one time step $t$ to the next. At each step, the death of a random node (Fig. \ref{fig_model}a, \textit{center}) yields an average decrease in probability of $-\frac{1}{N}P_t(k)$. On average, killing a node produces $\bar{k} = 2E/N$ disconnected edges that must be reattached. With probability $p$, each edge attaches preferentially (Fig. \ref{fig_model}a, \textit{bottom left}), connecting to a node of degree $k$ with probability $\frac{k}{2E}$; on average, this preferential attachment yields an increase in probability of $\bar{k}p\frac{k-1}{2E}P_t(k-1)$ and a decrease of $-\bar{k}p\frac{k}{2E}P_t(k)$. Alternatively, with probability $1-p$, each disconnected edge reattaches randomly (Fig. \ref{fig_model}a, \textit{bottom right}), yielding an increase in probability of $\bar{k}(1-p)\frac{1}{N}P_t(k-1)$ and a decrease of $-\bar{k}(1-p)\frac{1}{N}P_t(k)$. Combining these contributions and simplifying, we arrive at the master equation,
\begin{equation}
\label{eq_master}
P_{t+1}(k) = P_t(k) + \frac{1}{N}\Big[-P_t(k) + p\Big((k-1)P_t(k-1) - kP_t(k)\Big) + \bar{k}(1-p)\Big(P_t(k-1) - P_t(k)\Big)\Big].
\end{equation}

We are now prepared to study the evolution of the degree distribution. To compare against the real networks (for which $N \gtrsim E$), we begin by randomly placing $E = 10^3$ edges among $N = 2\times 10^4$ nodes, for an average degree $\bar{k} = 0.1$. Running the dynamics with equal amounts of preferential and random attachment (such that $p = 0.5$), we find that the master equation [Eq. (\ref{eq_master})] provides a close approximation to simulations (Fig. \ref{fig_model}b). As the connections rearrange, the degree distribution, which is initially Poisson (Fig. \ref{fig_model}b, \textit{left}), quickly broadens (Fig. \ref{fig_model}b, \textit{center}). Eventually, the distribution develops a clear power law $P(k) \sim k^{-\gamma}$ in the high--degree limit $k \gg 1$, with a realistic exponent $\gamma = 3$ (Fig. \ref{fig_model}b, \textit{right}). We therefore find that scale--free structure emerges naturally from our simple dynamics (Fig. \ref{fig_model}a).

The emergence of scale--free structure leaves an imprint on network properties beyond just the degree distribution. Consider, for example, the size of the network $n$, which (for consistency with the real networks) is defined as the number of nodes with at least one connection. As the dynamics unfold, edges tend to collect around a small number of high--degree hubs, thus decreasing the size of the network $n$ (Fig. \ref{fig_model}c). These hubs comprise the heavy tail of the degree distribution. To quantify this heavy tail, rather than using the variance of the degrees (which diverges for power--law distributions with $\gamma \le 3$), we instead compute the \textit{heterogeneity} $\frac{1}{2}\left<|k_i - k_j|\right>/\left<k\right>$, which is normalized to lie between zero and one (where $\left<\cdot\right>$ represents an average over degrees $k \ge 1$ and $\left<|k_i - k_j|\right>$ measures the average absolute difference in degrees).\cite{Lynn-07} As the network evolves, and scale--free structure emerges (Fig. \ref{fig_model}b), we see that the degree heterogeneity increases (Fig. \ref{fig_model}d). Notably, both the network size and degree heterogeneity approach steady--state values, with larger proportions $p$ of preferential attachment yielding networks that are smaller (Fig. \ref{fig_model}c), yet more heterogeneous (Fig. \ref{fig_model}d).

\noindent {\large \myfont Steady--state scale--free structure}

\noindent Thus far, we have explored the network dynamics numerically (using the master equation) and through simulations. To make analytic progress, we must solve for the steady--state degree distribution. Setting $P_t(k) = P_{t+1}(k) = P(k)$, the master equation reduces to the recursion relation
\begin{equation}
\label{eq_rec}
P(k) = \frac{p(k-1) + \bar{k}(1-p)}{1 + pk + \bar{k}(1-p)}P(k-1).
\end{equation}
In the thermodynamic limit $N,E\rightarrow \infty$ (holding fixed the average degree $\bar{k} = 2E/N$), one can then solve for the steady--state distribution
\begin{equation}
\label{eq_dist}
P(k) = \frac{1}{C}\frac{\Gamma \big(k + \bar{k}(\frac{1}{p} - 1)\big)}{\Gamma \big(k + \bar{k}(\frac{1}{p} - 1) + 1 + \frac{1}{p}\big)},
\end{equation}
where $C$ is the normalization constant and $\Gamma(\cdot)$ is Euler's gamma function. In what follows, we normalize $P(k)$ to run over positive degrees $k \ge 1$, such that $C = \frac{\Gamma(\frac{1}{p})\Gamma(1 + \bar{k}(\frac{1}{p}-1))}{\Gamma(1+\frac{1}{p})\Gamma(1 + \frac{1}{p} + \bar{k}(\frac{1}{p}-1))}$. In the high--degree limit $k \gg \bar{k}/p$, the above distribution falls off as a power law $P(k) \sim k^{-\gamma}$ with scale--free exponent $\gamma = 1 + \frac{1}{p}$ (see Methods). We therefore find that the network dynamics produce a wide range of exponents $\gamma \ge 2$ observed in real--world systems. Moreover, this scale--free structure depends only on the proportion $p$ of preferential attachment (independent from the average degree $\bar{k}$).

\begin{figure}[t]
\centering
\includegraphics[width = .8\textwidth]{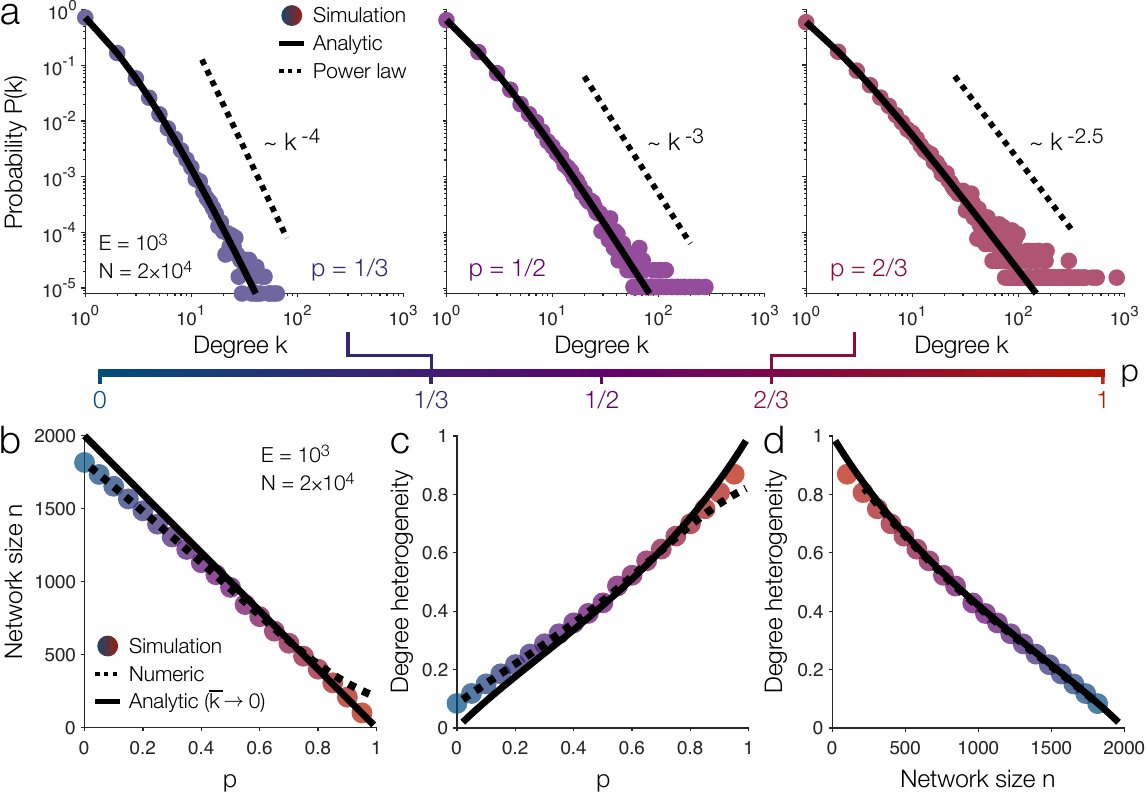} \\
\raggedright
\captionsetup{labelformat=empty}
{\spacing{1.25} \caption{\small \textbf{Fig.~\ref{fig_steady} $|$ Analytic predictions in steady--state.} \textbf{a}, Steady--state degree distributions for increasing proportions $p$ of preferential attachment in networks with $E = 10^3$ edges and $N = 2\times 10^4$ nodes (for average degree $\bar{k} = 0.1$). Data points depict simulations (see Methods), solid lines reflect the analytic prediction in Eq. (\ref{eq_dist}), and dashed lines illustrate power laws with the predicted exponents $\gamma = 1 + \frac{1}{p}$. \textbf{b-c}, Network size $n$ (\textbf{b}) and degree heterogeneity (\textbf{c}) as functions of $p$. \textbf{d}, Degree heterogeneity versus network size while sweeping over $p$. In panels \textbf{b}-\textbf{d}, data points are computed using simulations, dashed lines are calculated numerically using Eq. (\ref{eq_dist}), and solid lines are analytic predictions in the limit of sparse connectivity $\bar{k} \rightarrow 0$ (see Methods). \label{fig_steady}}}
\end{figure}

We confirm the analytic distribution [Eq. (\ref{eq_dist})] and the power--law tail in simulations (Fig. \ref{fig_steady}a). For equal amounts of preferential and random attachment ($p = 0.5$), the model generates a scale--free exponent $\gamma = 3$ (Fig. \ref{fig_steady}a, \textit{center}), as observed previously in Fig. \ref{fig_model}b. For larger proportions $p$ of preferential attachment, high--degree hubs become more prevalent, strengthening the heavy tail in $P(k)$ and decreasing the exponent $\gamma$ (Fig. \ref{fig_steady}a, \textit{right}). Indeed, as $p$ increases, the dynamics produce networks that are smaller (Fig. \ref{fig_steady}b) and more heterogeneous (Fig. \ref{fig_steady}c; see Methods for analytic predictions). Our model thus predicts a specific inverse relationship between network size and heterogeneity (Fig. \ref{fig_steady}d), which we will be able to test in real networks. Together, these results establish analytically that our simple network dynamics give rise to scale--free structure with realistic exponents $\gamma$.

\noindent {\large \myfont Modeling real networks}

\begin{figure}[t]
\centering
\includegraphics[width = .9\textwidth]{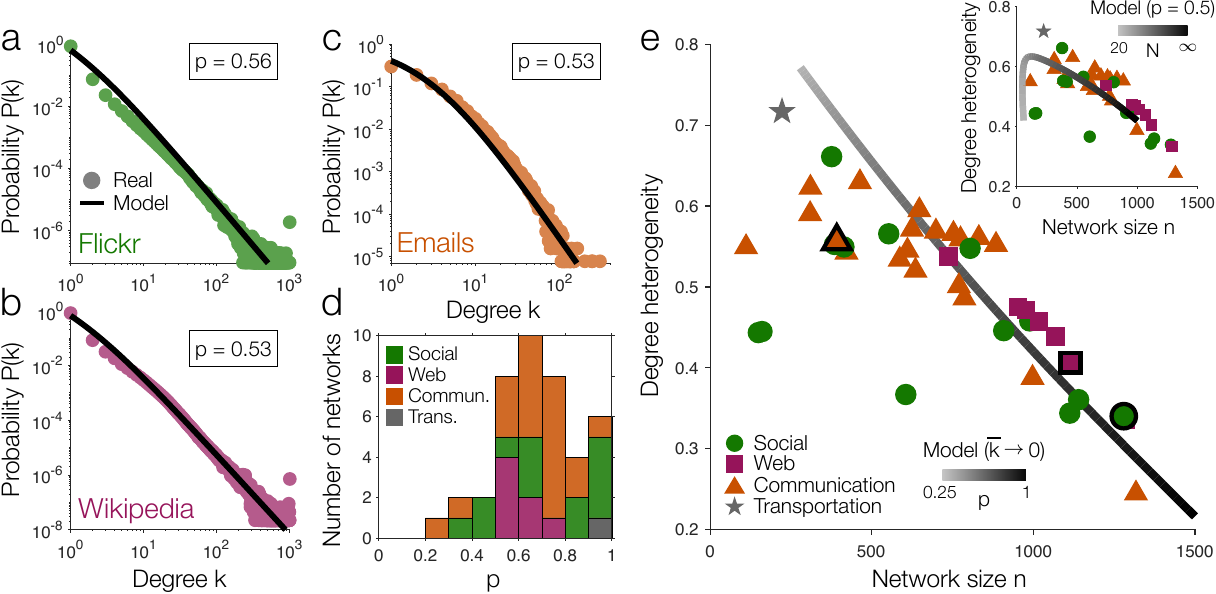} \\
\raggedright
\captionsetup{labelformat=empty}
{\spacing{1.25} \caption{\small \textbf{Fig.~\ref{fig_comp} $|$ Comparing real and model networks.} \textbf{a-c}, Analytic degree distribution in Eq. (\ref{eq_dist}) (solid lines) with $E = 10^3$, $N$ set to the number of nodes in a given dataset, and $p$ fit to the observed degree distributions (data points) for friendships on Flickr (\textbf{a}),\cite{Mislove-02} hyperlinks on English Wikipedia (\textbf{b}),\cite{Mislove-01} and emails between researchers (\textbf{c}).\cite{Paranjape-01} \textbf{d}, Distribution of preferential attachment proportions $p$ fit to the degree distributions of the real networks in Table \ref{tab} (see Supplementary Information for fits). Colors correspond to different network types. \textbf{e}, Degree heterogeneity as a function of network size $n$. Data points represent averages over different snapshots for individual real networks (Table \ref{tab}), and lines reflect numeric model predictions while sweeping over $p$ (with $\bar{k}\rightarrow 0$) or sweeping over the number of nodes $N$ (with $p = 0.5$; \textit{inset}). The networks in panels \textbf{a}-\textbf{c} are outlined in black. \label{fig_comp}}}
\end{figure}

\noindent Ultimately, we would like to use our model to study real--world systems. To compare against real networks (such as those in Fig. \ref{fig_real}), we fix the number of edges (here, $E = 10^3$) and approximate the number of nodes in the model $N$ by the total number that appear in a given dataset. This leaves one free parameter, the proportion $p$ of preferential attachment, which we can fit to the degree distribution of a given network (see Methods). For example, the networks in Fig. \ref{fig_real} are best described as arising from nearly equal amounts of preferential and random attachment ($p \approx 0.5$; Fig. \ref{fig_comp}a-c). In fact, despite only fitting one parameter, our simple model provides a surprisingly good description of nearly all the networks in Table \ref{tab} (see Supplementary Information)---even those that are merely heavy--tailed and not obviously scale--free (such as the emails in Fig. \ref{fig_comp}c). Across these networks, the proportion of preferential attachment ranges from 20\% to 100\%, slightly outpacing random attachment on average (Fig. \ref{fig_comp}d).

As we sweep over $p$, adjusting the ratio of preferential to random attachment, the model predicts a specific tradeoff between the size of a network and its degree heterogeneity (Fig. \ref{fig_steady}d). Computing the average properties (over different snapshots) for each of the real networks (Table \ref{tab}), we find a similar inverse relationship between network size and heterogeneity (Fig. \ref{fig_comp}e). If we instead hold $p$ fixed and sweep over the number of nodes $N$, the model also predicts the drop in degree heterogeneity observed in small networks (Fig. \ref{fig_comp}e, \textit{inset}). Moreover, even at the level of individual networks, we discover similar tradeoffs between size and heterogeneity across different snapshots (see Supplementary Information). We therefore find that our model not only captures the degree distributions observed in real--world systems (Fig. \ref{fig_comp}a-c; Supplementary Information), but it also predicts the relationships between different network properties (Fig. \ref{fig_comp}e).

\noindent {\large \myfont Extensions and robustness}

\begin{figure}[t!]
\centering
\includegraphics[width = \textwidth]{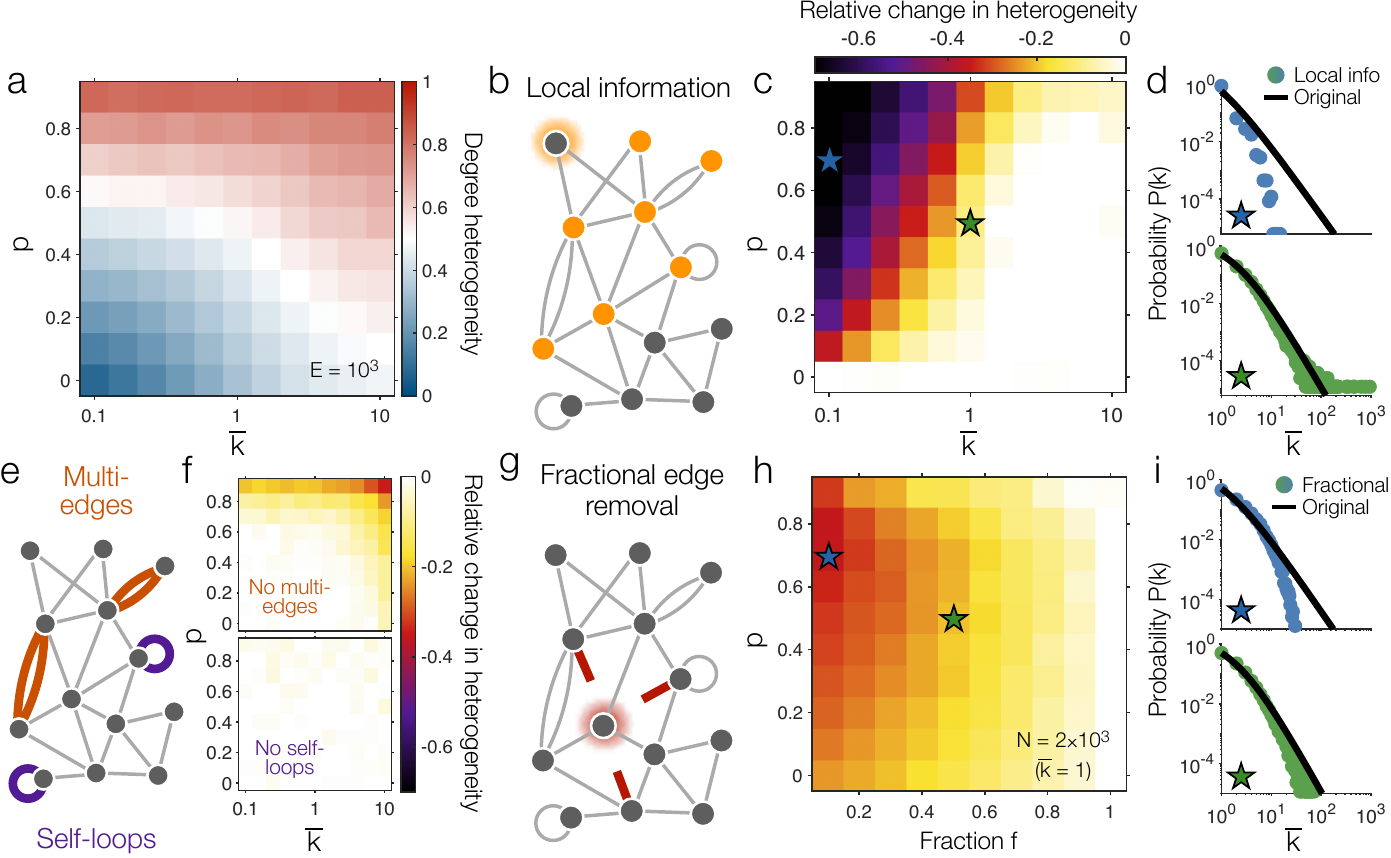} \\
\raggedright
\captionsetup{labelformat=empty}
{\spacing{1.25} \caption{\small \textbf{Fig.~\ref{fig_extend} $|$ Extending the original model.} \textbf{a}, Degree heterogeneity of the original model (Fig. \ref{fig_model}a) while sweeping over the preferential attachment proportion $p$ and average degree $\bar{k}$ for networks with $E = 10^3$ edges. \textbf{b}, Constraining to local information, each node can preferentially attach only to its neighbors and their neighbors. \textbf{c}, Relative change in degree heterogeneity after restricting to local information while sweeping over $p$ and $\bar{k}$. \textbf{d}, Degree distributions for the original model [Eq. (\ref{eq_dist}); solid lines] and with local information (data points) for parameters $p$ and $\bar{k}$ indicated in panel \textbf{c}. \textbf{e}, Multi--edges and self--loops are allowed in the original model. \textbf{f}, Relative change in heterogeneity when disallowing multi--edges (\textit{top}) or self--loops (\textit{bottom}). \textbf{g}, When a node dies, rather than removing all of its edges, one could remove only a fraction $f$. \textbf{h}, Relative change in heterogeneity with fractional edge removal while sweeping over $p$ and $f$ (for networks of average degree $\bar{k} = 1$). \textbf{i}, Degree distributions for the original model [Eq. (\ref{eq_dist}); solid lines] and with fractional edge removal (data points) for parameters $p$ and $f$ indicated in panel \textbf{h}. In all panels, values are computed using simulated networks with $E = 10^3$ edges (see Methods). \label{fig_extend}}}
\end{figure}

\noindent In designing the model (Fig. \ref{fig_model}a), we sought the simplest dynamics that would self--organize to produce scale--free structure. Given this simplicity, there are a number of natural extensions one could explore. To investigate the impact of model extensions on the degree distribution $P(k)$, we again consider the heterogeneity of degrees. In the original model (with the number of edges $E$ held fixed), as we sweep over the proportion $p$ of preferential attachment and the number of nodes $N$ (or, equivalently, the average degree $\bar{k} = 2E/N$), we arrive at a phase diagram for the network structure (Fig. \ref{fig_extend}a). As $p$ and $\bar{k}$ increase, the network dynamics produce degree distributions with heavier tails, thus increasing the degree heterogeneity (Fig. \ref{fig_extend}a).

When performing preferential attachment, we note that these simple dynamics (Fig. \ref{fig_model}a) rely on global information about the degrees of all the nodes in a network. In some scenarios, however, a node may only have access to local information about the degrees of nodes in its own neighborhood (for example, its neighbors and their neighbors; Fig. \ref{fig_extend}b).\cite{Wang-02, Pan-01} Restricting to local information, we find that the degree heterogeneity is significantly reduced for large $p$ (when preferential attachment dominates) and small $\bar{k}$ (when connections are sparse, and therefore local information becomes severely restrictive; Fig. \ref{fig_extend}c and d, \textit{top}). By contrast, for $\bar{k} \gtrsim 1$, the networks are dense enough that local information is sufficient to generate heterogeneous degrees (Fig. \ref{fig_extend}c) and, indeed, scale--free structure (Fig. \ref{fig_extend}d, \textit{bottom}).

Beyond global information, the original model also allows multi--edges (where two nodes are connected by multiple edges; Fig. \ref{fig_extend}e, \textit{top}) and self--loops (where a node connects to itself; Fig. \ref{fig_extend}e, \textit{bottom}). If we disallow multi--edges, the network dynamics still produce scale--free structure for all of parameter space besides $p \ge 0.9$ and $\bar{k} \gg 1$ (when networks are both highly heterogeneous and dense; Fig. \ref{fig_model}f, \textit{top}). Similarly, if we disallow self--loops, the degree distribution is almost entirely unaffected (Fig. \ref{fig_model}f, \textit{bottom}). As a final extension, when a node dies, rather than losing all of its connections, one could imagine that it only loses some fraction $f$ (Fig. \ref{fig_extend}g). In the limit $f = 0$, the dynamics halt and the network becomes static. As $f$ increases, so too does the degree heterogeneity, until at $f = 1$, we recover the original model (Fig. \ref{fig_extend}h). Indeed, as long as dying nodes lose a fraction $f \gtrsim 0.5$ of their edges, the model still produces power--law degree distributions (Fig. \ref{fig_extend}i, \textit{bottom}), which we confirm for different average degrees $\bar{k}$ (see Supplementary Information). In combination, these results demonstrate specific ways that the network dynamics can be extended, restricted, and generalized, while still giving rise to scale--free structure.

 \section*{\large Discussion}
\vspace{-30pt}
\noindent\rule{\textwidth}{.5pt}

\noindent Understanding how scale--free structure arises from fine--scale mechanisms is central to the study of complex systems. However, existing mechanisms typically require constant growth, an assumption that fails dramatically in many real--world networks. Even in networks that are usually viewed as growing, we show that individual snapshots (which cannot grow without bound by definition) can still exhibit scale--free structure (Fig. \ref{fig_real}). Here, we propose a simple model in which scale--free structure emerges naturally through the self--organization of nodes and edges. By allowing nodes to die, and letting connections rearrange under a mixture of preferential and random attachment, we show (both analytically and through simulations) that the degree distribution develops a power--law tail $P(k) \sim k^{-\gamma}$ (Fig. \ref{fig_model}). Moreover, the scale--free exponent (which takes the simple form $\gamma = 1 + \frac{1}{p}$) only depends on the proportion $p$ of preferential attachment and captures a wide range of values $\gamma \ge 2$ observed in real systems (Fig. \ref{fig_steady}). In fact, despite containing only one free parameter, the model provides a surprisingly good description of many real--world networks (Fig. \ref{fig_comp}; Supplementary Information). 

Given the simplicity of the model, one can immediately begin to extend the network dynamics to include additional features and mechanisms. For example, here we investigate the effects of local information, multi--edges, self--loops, and fractional edge removal (Fig. \ref{fig_extend}). Future work can build upon this progress to develop new models for the emergence of scale--free networks. Beyond node degrees, we note that power--law distributions also arise in many other contexts, from the strengths of connections in the brain and the frequencies of words in language to the populations of cities and the net worths of individuals.\cite{Clauset-01, Lynn-13} Do these power laws rely on the constant growth of a system? Or, instead, can scale--free distributions arise through the self--organization of existing resources? The framework presented here may provide fundamental insights to these questions.

\newpage

\section*{\large Methods}
\vspace{-30pt}
\noindent\rule{\textwidth}{.5pt}

\begin{methods}

\subsection{Simulations.} In each simulation (Figs. \ref{fig_model}, \ref{fig_steady}, and \ref{fig_extend}), we begin with $E = 10^3$ edges (to match the real networks) placed randomly among $N$ nodes (allowing multi--edges and self--loops) and a preferential attachment proportion $p$. Since the real networks tend to have average degrees $\bar{k} \lesssim 1$ (Table \ref{tab}), we consider $N = 2\times 10^4$ nodes for an average degree $\bar{k} = 0.1$ (unless otherwise specified). When simulating the dynamics, each step consists of one pass through the update rules in Fig. \ref{fig_model}a. Specifically, we first select a random node and remove all of its connections (Fig. \ref{fig_model}a, \textit{center}). Each disconnected edge is then reattached in one of two ways: (i) with probability $p$, the edge attaches preferentially (that is, to node $i$ with probability $\frac{k_i}{2E}$; Fig. \ref{fig_model}a, \textit{bottom left}), or (ii) with probability $1-p$, the edge attaches randomly (that is, to a node selected uniformly at random; Fig. \ref{fig_model}a, \textit{bottom right}). After repeating this process $N$ times, each node has died once (on average), which we refer to as one network update (Fig. \ref{fig_model}). 

When computing steady--state properties (as in Figs. \ref{fig_steady} and \ref{fig_extend}), we first perform 50 network updates of burn--in, which is sufficient for the network to reach steady--state (Fig. \ref{fig_model}). We then record 100 network samples, each interspersed by one network update. In Fig. \ref{fig_extend}, we investigate a number of model extensions, each of which can be implemented with minor changes to the above dynamics. To restrict to local information (Fig. \ref{fig_extend}b), when performing preferential attachment, each node is only allowed to connect to its neighbors and their neighbors. To remove multi--edges (Fig. \ref{fig_extend}e, \textit{top}), when performing preferential or random attachment, nodes are disallowed from connecting to their neighbors. Similarly, to remove self--loops (Fig. \ref{fig_extend}e, \textit{bottom}), nodes are disallowed from connecting to themselves. Finally, to implement fractional edge removal (Fig. \ref{fig_extend}g), when a node dies, it only loses a specified fraction $f$ of its connections.

\subsection{Analytic predictions.} Here, we derive a number of analytical results regarding the steady--state properties of the model (Fig. \ref{fig_steady}). As discussed above, beginning with the master equation [Eq. (\ref{eq_master})], one can solve for the steady--state degree distribution $P(k)$ [Eq. (\ref{eq_dist})] in the thermodynamic limit $N,E\rightarrow \infty$ with the average degree $\bar{k} = 2E/N$ held fixed. We remark that Euler's gamma functions of the form $\Gamma(x + c)$ diverge as $x^c$ in the limit $x \gg c$. Thus, in the high--degree limit $k \gg \bar{k}/p$, the degree distribution [Eq. (\ref{eq_dist})] falls off as a power law,
\begin{equation}
P(k) \propto \frac{\Gamma \big(k + \bar{k}(\frac{1}{p} - 1)\big)}{\Gamma \big(k + \bar{k}(\frac{1}{p} - 1) + 1 + \frac{1}{p}\big)} \sim \frac{k^{\bar{k}(\frac{1}{p} - 1)}}{k^{\bar{k}(\frac{1}{p} - 1) + 1 + \frac{1}{p}}} = k^{-(1 + \frac{1}{p})}.
\end{equation}
We therefore find that the degree distribution has a power--law tail $P(k) \sim k^{-\gamma}$ with scale--free exponent $\gamma = 1 + \frac{1}{p}$. We note that $\gamma$ only depends on the proportion $p$ of preferential attachment (independent of the average degree $\bar{k}$), and can achieve any value $\gamma \ge 2$.

Using the degree distribution [Eq. (\ref{eq_dist})], we can derive analytic predictions for different network properties. For example, we note that the network size $n$ (that is, the number of nodes with degree $k \ge 1$) is given simply by $n = N(1 - P(0))$, where $P(0)$ is the probability of a node having no connections. Throughout the paper, we normalize $P(k)$ to run over the positive degrees $k \ge 1$, but if we allow it to run over all degrees $k \ge 0$, then the normalization constant takes the form $C = \frac{\Gamma(\frac{1}{p})\Gamma(\bar{k}(\frac{1}{p} - 1))}{\Gamma(1 + \frac{1}{p})\Gamma(\frac{1}{p} + \bar{k}(\frac{1}{p} - 1))}$. Solving for the zero--degree probability $P(0) = (1 + \bar{k}(1 - p))^{-1}$, we arrive at an analytic prediction for the network size,
\begin{equation}
n = \frac{N\bar{k}(1-p)}{1 + \bar{k}(1-p)} = \frac{2E(1-p)}{1 + \bar{k}(1-p)}.
\end{equation}
Thus, in the limit of sparse connectivity $\bar{k}\rightarrow 0$, we find that $n = 2E(1 - p)$, as illustrated in Fig. \ref{fig_steady}b.

Let's now consider the degree heterogeneity $\frac{1}{2}\left<|k_i - k_j|\right>/\left<k\right>$, where $\left<\cdot\right>$ indicates an average over positive degrees $k \ge 1$. Using $P(k)$, one can compute $\left<k\right> = \bar{k} + \frac{1}{1-p}$, which diverges as $p \rightarrow 1$ (and as $\gamma \rightarrow 2$). In the sparse connectivity limit $\bar{k}\rightarrow 0$, we can compute the average absolute difference in degrees $\left<|k_i - k_j|\right>$ in a similar fashion, resulting in the following expression for the degree heterogeneity,
\begin{equation}
H = \frac{p}{1+p} \,{}_3 F_2\Big(1,1,2;\, 1 + \frac{1}{p}, 2 + \frac{1}{p};\, 1\Big),
\end{equation}
where ${}_r F_q(\cdot)$ is the generalized hypergeometric function. In the limit of purely random dynamics ($p\rightarrow 0$), we see that the heterogeneity vanishes. Conversely, in the limit of purely preferential attachment ($p \rightarrow 1$), we have ${}_3 F_2(1,1,2;\, 2, 3;\, 1) = 2$, and so the heterogeneity reaches its maximum possible value of one. This analytic prediction is illustrated in Fig. \ref{fig_steady}c.

\subsection{Analyzing real--world networks.} We analyze 41 different networks listed in Table \ref{tab} (for individual descriptions and references, see Supplementary Information). Each network was selected based on two criteria: (i) temporal dynamics (in order to investigate the evolution of network structure) and (ii) a heavy--tailed (although not necessarily scale--free) degree distribution. These network fall into four distinct categories: social (reflecting the connections between people), web (representing hyperlinks between websites or the physical wiring of the Internet), communication (comprised of messages between people), and transportation (mapping flights between cities). All of the networks analyzed in this paper have been made openly available (see Data Availability).

\setcounter{figure}{0}
\begin{figure}
\textbf{Table \ref{tab} $|$ Real temporal networks and their properties} \\[.5em]
{\fontsize{9.3}{8}\selectfont \myfont
\begin{tabular}{l l l l l l c l}
\hline
\\[-1em]
{\footnotesize \textbf{Type}} & Name & $N$ & $T$ & $\bar{k}$ & $n$ & Deg. het. & $p$ \\[.1em]
\hline
\hline
\\[-1em]
{\footnotesize \textcolor{MyGreen}{\textbf{Social}}} & Flickr & 920,743 & 8,311,777 & 2.2$\times$10$^\text{-3}$ & 1,280 & 0.34 & 0.56 \\
& Twitter & 18,470 & 61,157 & 0.11 & 1,140 & 0.36 & 0.46 \\
& Facebook & 46,952 & 876,993 & 0.04 & 1,113 & 0.34 & 0.36 \\
& Wikipedia (elections) & 7,118 & 103,675 & 0.28 & 376 & 0.66 & 1 \\
& Wikipedia (conflict) & 116,836 & 2,917,785 & 0.02 & 552 & 0.57 & 0.85 \\
& Epinions & 56,779 & 262,376 & 0.04 & 908 & 0.45 & 0.60 \\
& Digg & 279,374 & 1,729,983 & 7.2$\times$10$^\text{-3}$ & 989 & 0.46 & 0.62  \\
& Loans & 89,269 & 3,394,979 & 0.02 & 804 & 0.55 & 0.60 \\
& DBLP & 1,824,701 & 14,743,872 & 1.1$\times$10$^\text{-3}$ & 605 & 0.37 & 0.43 \\
& Bitcoin (alpha) & 3,783 & 24,186 & 0.53 & 385 & 0.55 & 0.99 \\
& Bitcoin (OTC) & 5,881 & 35,592 & 0.34 & 414 & 0.55 & 0.87 \\
& Coauthors (HEP--Ph) & 16,959 & 2,322,259 & 0.12 & 160 & 0.44 & 0.98 \\
& Coauthors (HEP--Th) & 6,798 & 290,597 & 0.29 & 149 & 0.44 & 1 \\
\hline
\\[-1em]
{\footnotesize \textcolor{MyMagenta}{\textbf{Web}}} & Wikipedia (English) & 1,870,709 & 39,953,145 & 1.1$\times$10$^\text{-3}$ & 1,115 & 0.41 & 0.53 \\
& Wikipedia (French) & 2,212,682 & 41,724,673 & 9.0$\times$10$^\text{-4}$ & 1,017 & 0.46 & 0.60 \\
& Wikipedia (German) & 2,166,669 & 58,721,812 & 9.2$\times$10$^\text{-4}$ & 1,068 & 0.44 & 0.60 \\
& Wikipedia (Dutch) & 1,204,009 & 25,956,564 & 1.7$\times$10$^\text{-3}$ & 978 & 0.47 & 0.60 \\
& Wikipedia (Italian) & 1,039,252 & 15,341,526 & 1.9$\times$10$^\text{-3}$ & 953 & 0.47 & 0.60 \\
& Youtube & 2,234,127 & 7,787,826 & 9.0$\times$10$^\text{-4}$ & 1,287 & 0.33 & 0.60 \\
& Internet & 16,564 & 104,393 & 0.12 & 737 & 0.54 & 0.75 \\
\hline
\\[-1em]
{\footnotesize \textcolor{MyOrange}{\textbf{Communication}}} & Emails & 986 & 329,910 & 2.03 & 392 & 0.56 & 0.53 \\
& Wikipedia talk (English) & 2,987,535 & 24,981,162 & 6.7$\times$10$^\text{-4}$ & 770 & 0.50 & 0.63 \\
& Wikipedia talk (French) & 1,420,367 & 4,641,928 & 1.4$\times$10$^\text{-3}$ & 698 & 0.57 & 0.75 \\
& Wikipedia talk (German) & 519,403 & 6,729,793 & 3.9$\times$10$^\text{-3}$ & 586 & 0.53 & 0.64 \\
& Wikipedia talk (Spanish) & 497,446 & 2,702,879 & 4.0$\times$10$^\text{-3}$ & 624 & 0.57 & 0.70 \\
& Wikipedia talk (Dutch) & 225,749 & 1,554,699 & 8.9$\times$10$^\text{-3}$ & 463 & 0.63 & 0.78 \\
& Wikipedia talk (Italian) & 863,846 & 3,067,680 & 2.3$\times$10$^\text{-3}$ & 647 & 0.59 & 0.76 \\
& Wikipedia talk (Japanese) & 397,635 & 1,031,378 & 5.0$\times$10$^\text{-3}$ & 753 & 0.57 & 0.77 \\
& Wikipedia talk (Chinese) & 1,219,241 & 2,284,546 & 1.16$\times$10$^\text{-3}$ & 831 & 0.56 & 0.81 \\
& Wikipedia talk (Arabic) & 1,095,799 & 1,913,103 & 1.8$\times$10$^\text{-3}$ & 884 & 0.55 & 0.84 \\
& Yahoo & 100,001 & 3,179,718 & 0.02 & 1318 & 0.24 & 0.34 \\
& Enron & 87,101 & 1,147,130 & 0.02 & 775 & 0.56 & 0.75 \\
& Slashdot & 51,068 & 140,715 & 0.04 & 787 & 0.49 & 0.62 \\
& Super User & 193,976 & 1,439,111 & 0.01 & 635 & 0.52 & 0.54 \\
& College & 1,899 & 59,835 & 1.05 & 310 & 0.62 & 1 \\
& Manufacturing & 167 & 82,927 & 11.98 & 111 & 0.55 & 0.54 \\
& Ubuntu & 159,313 & 964,417 & 0.01 & 610 & 0.54 & 0.61 \\
& Linux & 27,927 & 1,096,440 & 0.07 & 309 & 0.59 & 0.78 \\
& Stack Overflow & 2,601,977 & 63,497,050 & 7.7$\times$10$^\text{-4}$ & 997 & 0.39 & 0.28 \\
& Math Overflow & 24,818 & 506,550 & 0.08 & 422 & 0.54 & 0.63 \\
\hline
\\[-1em]
{\footnotesize \textcolor{MyGrey}{\textbf{Transportation}}} & Flights & 2,322 & 7,287,850 & 0.86 & 222 & 0.72 & 1 \\
\hline
\end{tabular}} \\[-.5 em]
\captionsetup{labelformat=empty}
{\spacing{1} \caption{\footnotesize For each network, we list its type, name, total number of nodes $N$, and total number of edges $T$. Dividing each network into snapshots of $E = 10^3$ edges each, we list the average degree $\bar{k} = 2E/N$, average network size $n$ (that is, the average number of nodes in each snapshot), the average degree heterogeneity, and the proportion $p$ of preferential attachment that best describes the degree distribution. For descriptions of the networks, distributions of the above quantities, and references, see Supplementary Information. \label{tab}}}
\end{figure}

In order to study the evolution of real--world systems, all of the networks chosen for analysis are temporal, each consisting of a list of edges $(i_t, j_t)$ with time stamps $t$. For each network, we denote the total number of nodes by $N$ and the total number of edges (or the length of the dataset) by $T$. When examining the temporal dynamics, we divide the edges $(i_t, j_t)$ into groups of size $E = 10^3$, the largest number that could be applied to all networks. Grouping the edges this way results in a sequence of $\lfloor{T/E}\rfloor$ network snapshots (Fig. \ref{fig_real}a). Averaging over these snapshots, we can compute quantities such as the network size $n$ (or the average number of nodes in each snapshot) and the degree heterogeneity (Fig. \ref{fig_comp}e).

To compute the degree distribution $P(k)$ for a given network, we count the number of nodes of degree $k$ across all snapshots (Fig. \ref{fig_real}c,e,g). By definition, each degree distribution is normalized to run over all positive degrees $k \ge 1$. To estimate the proportion of preferential attachment $p$ that best describes a given network, we fit the analytic distribution in Eq. (\ref{eq_dist}) to the network's measured degree distribution. Specifically, we begin by setting $E = 10^3$ and letting $N$ (the total number of nodes in our model) equal the total number of nodes in a given dataset, which results in an estimate for the average degree $\bar{k} = 2E/N$ (see Table \ref{tab}). This leaves one free parameter $p$, which we compute by minimizing the root--mean--square error of the log probabilities. The degree distributions and model fits for all of the networks are displayed and discussed in the Supplementary Information. The gradient descent algorithm used to compute $p$ has been made openly available (see Code Availability).

\end{methods}

\section*{Data Availability}

The networks analyzed in this paper are openly available at \\ \texttt{github.com/ChrisWLynn/Emergent{\_}scale{\_}free}.

\section*{Code Availability}

The code used to perform the analyses in this paper is openly available at \\ \texttt{github.com/ChrisWLynn/Emergent{\_}scale{\_}free}.

%% Put the bibliography here, most people will use BiBTeX in
%% which case the environment below should be replaced with
%% the \bibliography{} command.

\newpage

%\section*{References}
\section*{\large References}
\vspace{-30pt}
\noindent\rule{\textwidth}{.5pt}

\bibliographystyle{naturemag}
\bibliography{ScaleFreeBib}

%% Here is the endmatter stuff: Supplementary Info, etc.
%% Use \item's to separate, default label is "Acknowledgements"
\newpage
\begin{addendum}

\item[Supplementary Information.] Supplementary text and figures accompany this paper.

\item[Acknowledgements.] This work was supported in part by the National Science Foundation, through the Center for the Physics of Biological Function (PHY--1734030) and a Graduate Research Fellowship (C.M.H.); by the James S. McDonnell Foundation through a Postdoctoral Fellowship Award (C.W.L.); and by the National Institutes of Health BRAIN initiative (R01EB026943).

\item[Citation diversity statement.] Recent work in several fields of science,\cite{Mitchell-01, Dion-01, Caplar-01, Dworkin-01, Teich-01, Bertolero-01} has identified citation bias negatively impacting women and other minorities. Here we sought to proactively consider choosing references that reflect the diversity of the field in thought, form of contribution, gender, and other factors. Excluding (including) self--citations to the current authors, our references contain 39\% (37\%) women lead authors and 24\% (33\%) women senior authors.
 
\item[Author Contributions.] C.W.L. conceived the project, designed the model, and performed the analysis with input from C.M.H. and S.E.P. C.W.L. wrote the manuscript and Supplementary Information; and C.M.H. and S.E.P. edited the manuscript and Supplementary Information.
 
\item[Competing Interests.] The authors declare no competing financial interests.
 
\item[Corresponding Author.] Correspondence and requests for materials should be addressed to C.W.L. \\(cwlynn@princeton.edu).
 
\end{addendum}

\end{document}